\documentclass[pre,aps,reprint,amsmath,amssymb,longbibliography]{revtex4-2}

\usepackage{graphicx}% Include figure files
\usepackage{dcolumn}% Align table columns on decimal point
\usepackage{bm}% bold math
\usepackage{physics}
\usepackage{natbib}
\usepackage{amsmath,amssymb}
\usepackage{hyperref}
\hypersetup{colorlinks = true,citecolor = blue,urlcolor = black}
\begin{document}

\title{Wrinkling and imaging of thin curved sheets}

\author{Megha Emerse}
\affiliation{School of Science and Technology, Nottingham Trent University, Nottingham NG11 8NS, UK}
\author{Lucas Goehring}
\email{E-mail: lucas.goehring@ntu.ac.uk}
\affiliation{School of Science and Technology, Nottingham Trent University, Nottingham NG11 8NS, UK}

\date{\today}

\begin{abstract}
Thin films or sheets subjected to external forces often undergo mechanical instability, leading to regular patterns of wrinkles, folds, and creases.  As can be anticipated from the difficulty of flattening a curved globe, any natural curvature of the sheet will have a strong influence on these instabilities. Here, we develop a non-invasive synthetic schlieren imaging technique to image and reconstruct the surface of wrinkling curved sheets, confined to float on water. Our method circumvents the small-amplitude limit of related imaging techniques, and we demonstrate robust means to estimate the reconstruction accuracy.  We then evaluate how the sign and magnitude of Gaussian curvature affects the wrinkling of thin curved sheets, and compare observations of the wrinkle wavelength, amplitude and domain structure with recent theoretical predictions.  While generally validating model predictions, we find that the assumption of a conserved surface area during wrinkling should be treated with some care.   The control of wrinkling behavior demonstrated here can have application in the design of liquid lenses, microfluidics, active textured surfaces, and flexible electronic components.
\end{abstract}

\maketitle

\section{Introduction}

\begin{figure}[t]
\includegraphics[width=\linewidth]
{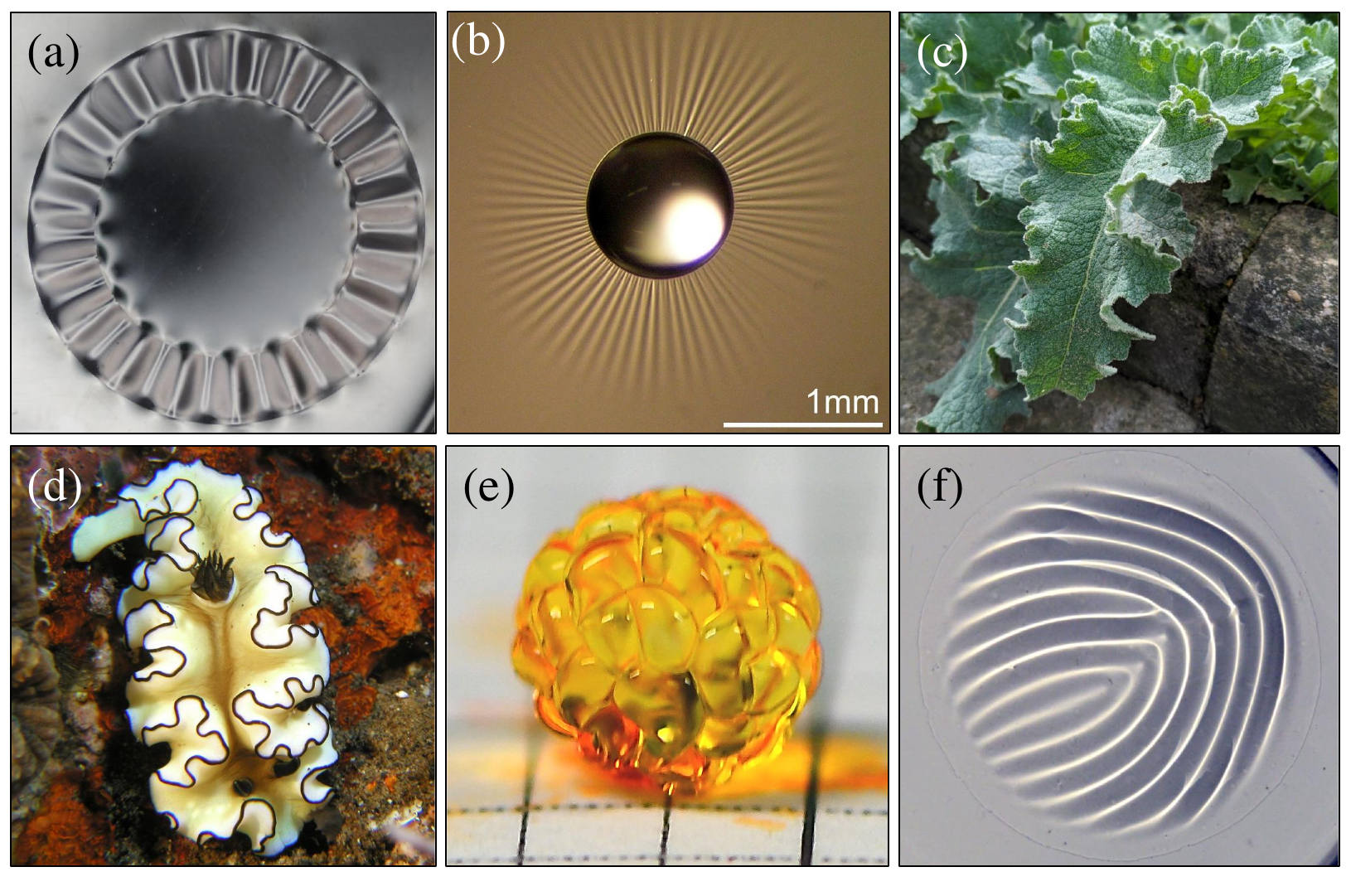}
\caption{\label{fig:1} Patterns due to mechanical instabilities like wrinkling, buckling and folding arise frequently in nature.  Examples include: thin sheets floating on water and forced by (a) surface tension differences across an annulus~\cite{pineirua2013capillary,paulsen2017geometry} or (b) the weight and surface tension of a droplet~\cite{huang2007capillary}; crinkling of the edges of (c) a leaf~\cite{sharon2002,Marder2003,sharon2004leaves} or (d) sea slug~\cite{sharon2004leaves}; (e) creasing observed during the swelling of hydrogel beads~\cite{bertrand2016dynamics}, reminiscent of cortical folding~\cite{jalil2015cortical,tallinen2016}; and (f) wrinkling of a thin spherical shell constrained to float on a water bath~\cite{Aharoni2017,Albarran2018curvature,Tobasco2022}. Images are reproduced or adapted from (a)~\cite{pineirua2013capillary}, (b)~\cite{huang2007capillary}, (d)~\cite{Watanabe2005} and (f)~\cite{Albarran2018curvature}, with permission.} 
\end{figure}

When a thin sheet is confined beyond some critical point, such as by being compressed along its edges, or as the result of differential growth, it responds by forming wrinkles instead of deforming uniformly~\cite{bowden1998spontaneous,cerda2003geometry,sharon2004leaves,huang2007capillary,diamant2011,li2012mechanics,paulsen2019,wang2022mechanics}.  
This behavior redistributes stress and relaxes excessive strain energy, through the bending of the sheet.  The results include a remarkable range of patterns, such as those shown in Fig.~\ref{fig:1}.   As an example, the relatively simple case of the wrinkling of a flat bi-layer, such as a floating membrane~\cite{audoly2011,diamant2011} or a coating on a thick substrate~\cite{bowden1998spontaneous,mahadevan2005}, is governed by two primary factors. The material's bending resistance, or stiffness, tends to promote wrinkles with larger wavelengths, since a stiffer material better resists bending over short distances. In addition, a restoring force acts to return the sheet to its unwrinkled state by counteracting the energetic cost of deforming the sheet's substrate. This force discourages large-amplitude deviations of the film's shape from its reference configuration. The equilibrium between these influences determines the characteristic size and pattern of the wrinkles that emerge on thin sheets in diverse scenarios~\cite{danov2010elastic,vella2011wrinkling,Finn2019,tan2020bioinspired}.  An elegant demonstration of this involves a flexible annular sheet, floating on water, as in Fig.~\ref{fig:1}(a).  By adding surfactant outside the annulus, a difference in surface tension can be set up between its outer and inner perimeters.  The film is then effectively placed into compression, which results first in the regular radial wrinkling~\cite{pineirua2013capillary} and then the buckling~\cite{paulsen2017geometry} of the film, beyond some critical differences in the surface tension.  By placing a droplet of water on a film, as in Fig.~\ref{fig:1}(b), the pull provided by the droplet's surface tension acting on the film, along with the indentation force of its own weight, can lead to a similar wrinkling response~\cite{huang2007capillary}.

More generally, surfaces need not be flat.  For example, the differential, volumetric growth of tissues can affect the morphological stability of their variously curved surfaces or interfaces. As in Fig.~\ref{fig:1}(c--d), this can be observed in the crinkles on the edges of leaves, flowers, sea slugs, or torn plastic bags~\cite{sharon2002,Marder2003,sharon2004leaves}, as well as in the folds of organs like the brain~\cite{jalil2015cortical,tallinen2016} or intestines~\cite{Shyer2013}.  Surface curvature is also an important factor for designing wrinkle patterns and transitions into materials and structures, particularly in curved film/substrate and core-shell geometries~\cite{li2011coreshell,bertrand2016dynamics,Paulsen2016}, like the case shown in Fig.~\ref{fig:1}(e).  
The curvature anywhere on a surface can be defined by two principal curvatures, $\kappa_1$ and $\kappa_2$, which give its maximum and minimum curvatures, with respect to a normal or tangent plane.   The Gaussian curvature is their product, $\kappa=\kappa_1\kappa_2$, while the reciprocals of the principal curvatures define the extreme values of the radius of curvature of circles that fit to the surface at that point, $R_1$ and $R_2$. Since curvature can vary locally, it provides the opportunity for a high level of control over wrinkle patterns.  

\begin{figure}[t]
\includegraphics[width=\linewidth]{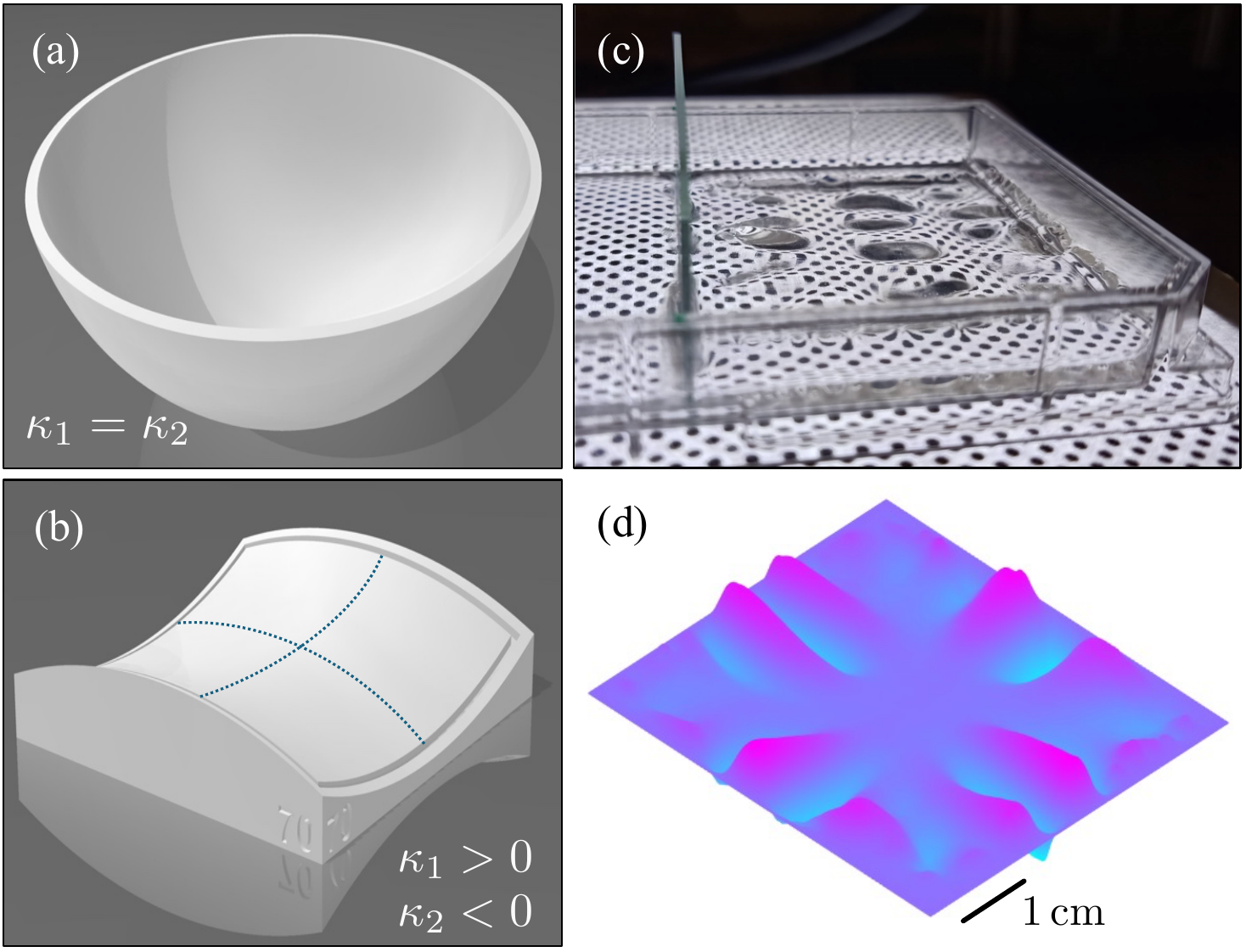}
\caption{\label{fig:2} Thin shells tend to wrinkle, when confined by a flat substrate. We made shells with intrinsic (a) positive and (b) negative Gaussian curvature, $\kappa = \kappa_1\kappa_2$, by casting PDMS films onto appropriately curved molds.  (c) Once cured, the films are peeled off and floated on the surface of a water bath, where they adopt a wrinkled shape.  (d) By observing the distortion of features projected through the film from below, a three-dimensional surface profile of the film is reconstructed.}
\end{figure}

Thin floating shells are beginning to be used to explore the wrinkling of curved sheets in more detail~\cite{Aharoni2017,Albarran2018curvature,Tobasco2021,Tobasco2022}, as in Fig.~\ref{fig:1}(f).  The mismatch between the inherent curvature of such a film, and the fluid surface on which it is supported, gives rise to a geometric constraint that is equivalent to lateral confinement~\cite{Albarran2018curvature,Tobasco2022}.  Surprisingly, even a small amount of curvature can easily dominate over surface tension effects in the level of confinement that it represents (e.g. see Fig.~S2 of Ref.~\cite{Aharoni2017}).  This system has also been shown to be particularly rich in patterns, with studies exploring the effects of the size and shape of a piece cut out of a curved surface on the wrinkle patterns~\cite{Aharoni2017,Albarran2018curvature,Tobasco2022}, for example.  Here, we use the floating shell geometry to experimentally test the predictions of how curvature affects the wavelength, amplitude, and domain structure of wrinkles in thin curved sheets.  Films cut from spherical caps, as in Fig.~\ref{fig:2}(a) are used for the case of positive curvature, $\kappa>0$, while hyperboloids are used as the basis for surfaces with negative Gaussian curvature, $\kappa<0$, as in Fig.~\ref{fig:2}(b). 

Although floating sheets and shells offer a simple, well-defined system in which to study wrinkling, the accurate imaging of a floating shell is challenging.   Shadowgraph methods are frequently used~\cite{Aharoni2017,paulsen2017geometry,Albarran2018curvature}, involving light being focused or dispersed by passing through the film, but it is difficult to convert the resulting light intensities into film relief.  Cross-sectional profiles of wrinkling patterns can be more accurately measured by imaging their intersection with a laser sheet~\cite{Albarran2018curvature}, for example. As an alternative for full surface mapping, synthetic schlieren imaging has been developed to image the dynamic topography of liquid interfaces~\cite{moisy2009synthetic,wildeman2018real}, in the limits of small surface slopes and amplitudes.  This method is based on how an uneven fluid layer distorts the image of a fixed, underlying reference pattern, in the same way that ripples on a shallow pond will distort one's view of what lies beneath the water.  It can also be used to visualize density variations or refractive index gradients in transparent media like gases or liquids in real-time, aiding in the study of heat, shock waves and fluid flows~\cite{gomit2022free}, and has been applied to image wrinkles in floating annular films~\cite{pineirua2013capillary}, and to monitor the evaporation of droplets~\cite{Kilbride2023}.   

A recent review of free-surface imaging~\cite{gomit2022free} gives a detailed comparison of surface reconstruction methods, including synthetic schlieren techniques and other non-intrusive methods like fringe projection profilometry and 3D scanning systems, which can capture surface topographies but which face limitations due to complexity, cost, and adaptability to transparent surfaces or surfaces of low reflectivity.   Refraction-based methods, like synthetic schlieren, are noted for their high accuracy in slope measurement, and for simplicity in implementation~\cite{gomit2022free}.  

Here we introduce a simple ray-tracing based, non-invasive imaging method and use it to map out the surfaces of floating wrinkled films with different inherent curvatures, as demonstrated in Fig.~\ref{fig:2}(c,d).  This reformulation of synthetic schlieren imaging avoids the need for any small-amplitude approximation in the inferred surface height profile. 

To make thin shells with positive and negative Gaussian curvatures, we use hemispherical and saddle-shaped substrates, respectively, with a wide range of curvatures.  When these curved shells are floated on a water bath, we measure how the wrinkle wavelength, amplitude and domain structure vary with the sign and magnitude of the Gaussian curvature, $\kappa$.  The wrinkle wavelength is independent of curvature, as predicted from models built around the Euler elastica concept~\cite{Albarran2018curvature}, while the wrinkle amplitude scales with $\kappa^{1/2}$.  This follows from the idea that the wrinkles take up the excess area of the curved film, compared to its two-dimensional projection~\cite{Aharoni2017}, although stretching of the film leads to differences in how much excess area is available for the cases of positive and negative curvature, respectively.  The wrinkles themselves are divided into domains consisting of parallel features, separated by domain boundaries, with the domain shapes closely following recently elucidated rules of how reciprocal force diagrams can be used to analyze curved films~\cite{Tobasco2021,Tobasco2022}.   

\section{Materials and Methods}

\begin{figure*}[t]
\includegraphics[width=\linewidth]{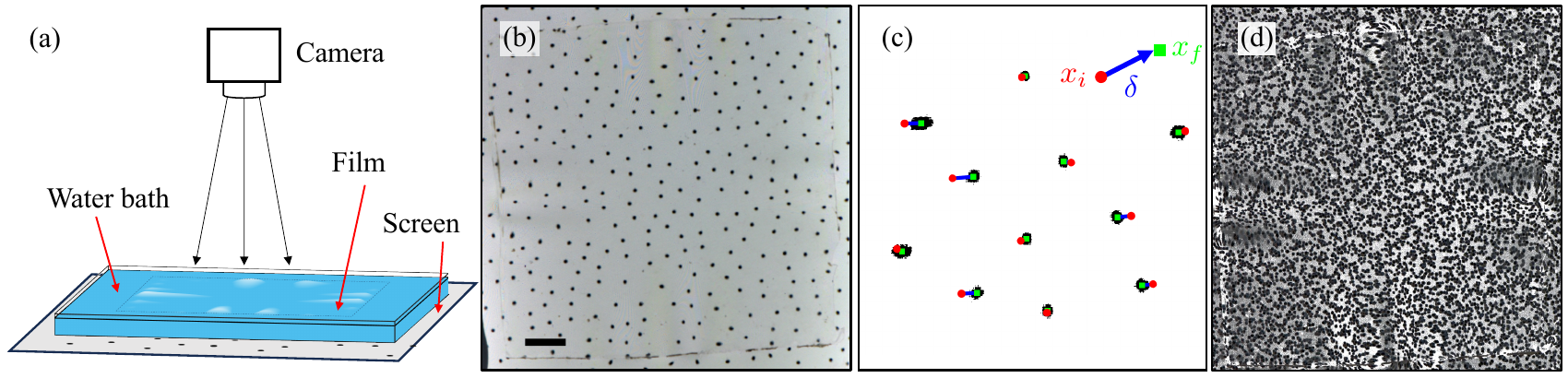}
\caption{\label{fig:3} Imaging methods. (a) The setup involves a digital screen projecting a pattern through a liquid surface, which is deformed by the presence of a floating, wrinkled film. (b) A typical image recorded on an overhead camera shows a random close-packed dot pattern, as seen through the film.  Here, the scale bar is 1 cm, and bandpass filtering is used to minimize the moir\'e effect from the periodicity of the display pixels. (c) A close-up view of a processed binary image shows black dots on a white background.  Matching up the positions of the centroids of the dot features before (red circles) and after (green squares) emplacing the film leads to a set of dot displacement vectors (blue lines). (d) By overlaying the results of many distinct patterns, a clear view of the distortion caused by the film emerges.}
\end{figure*}

\subsection{Thin shells}
\label{sec:Shells}

The methods of fabricating and floating curved shells is outlined in Fig.~\ref{fig:2}.  Shapes with surfaces of negative Gaussian curvature, $\kappa < 0$, were designed in CAD and fabricated on a Formlabs Form2 3D printer, with 100 $\mu$m vertical and 150 $\mu$m horizontal printing resolution, using a high-temperature V2 photopolymer resin.  A batch of ten saddle-shaped molds were made, each with a base area of 55$\times$55 mm$^2$, a positive radius of curvature,  $R_1$, varying from $40$ to $130$ mm, and a fixed negative radius of curvature of $R_2 = -70$ mm.  High-impact polystyrene sheets were vacuum thermo-formed to these surfaces to create the substrates on which films were subsequently cast.  For surfaces with a positive Gaussian curvature, acrylic hemispherical shells were purchased from Uhat DIY. The radius of curvature of these substrates varied from $30$ to $70$ mm, with $R_1 = R_2$. 

Polydimethylsiloxane (PDMS) was prepared by mixing Sylgard 184 (Dow) base to curing agent in a 10:1 ratio, and degassing under vacuum.  Films were made by drop casting PDMS onto the thermo-formed substrates, adapting methods from Ref.~\cite{Albarran2018curvature}.  For this, the films were inverted ($\kappa > 0$, see methods in Ref.~\cite{lee2016fabrication}) or steeply inclined ($\kappa < 0$), allowing the excess PDMS to drain and leave an even layer.  Films of thickness of $t\simeq 100\, \mu$m were made by repeatedly coating a substrate, and then curing at 80$^\circ$C for 4 hours.  To aid in the removal of the polymer films, a layer of PVA glue was deposited and dried on the substrates, prior to the drop casting.  This layer was removed by gentle agitation in water, after the cured film had been peeled off its mold.  Once removed, films were trimmed to a square cross section with edge lengths of $W = 50$ mm.  

The thickness $t$ of the prepared PDMS films was measured using optical coherence tomography (OCT)~\cite{manukyan2013imaging, cheung2015ultra} assuming the refractive index of the PDMS to be 1.40~\cite{kacik2014measurement}.  The films used here had thicknesses between 100 and 118 $\mu$m.  Five measurements of $t$ were made for each film, including at points in all four quadrants and at the center of the film.  The film thickness was uniform to within 10-20$\%$, with an average standard deviation in $t$ of 17$\%$ (19 $\mu$m). The Young's modulus of the PDMS was measured through tensile testing (MultiTest 2.5-dV, Mecmesin).  Three samples with a 110$\times$10$\times$5.7 mm$^3$ dog-bone geometry were tested to 10$\%$ extension, with a Young's modulus of $E = 2.2 \pm 0.3$ MPa measured by a linear least-squares fit to the stress-strain data. The Poisson ratio of the PDMS was taken to be $\nu = 0.5$~\cite{muller2019quick,hemmerle2021measuring}.

\subsection{Image acquisition and processing}
\label{sec:Image processing}

Figure~\ref{fig:3} gives an overview of our imaging methods.  In each experiment a transparent, plastic Petri dish was filled with deionized water to a depth of $h_0= 5$ mm.  A curved film was then gently set to float on the surface of the water bath, where it adopted a wrinkled shape.

The floating shells were imaged by projecting patterns of dots through the films, from below, as sketched in Fig.~\ref{fig:3}(a).  As projections, random close-packed patterns~\cite{scott1969density,baranau2014random} were generated by placing dots at random onto a two-dimensional space, while enforcing a minimum separation of 0.36 cm between the dots.  A series of 30 such patterns, each with dots at an average density of $\rho=$ 5 cm$^{-2}$, was used for each experiment.  A tablet computer was used as a projecting screen, and an overhead digital camera (Nikon D5100) collected images.  All experiments were done under calm conditions, with no other users in the lab, and using computer controls to trigger the imaging sequence.  Otherwise, a ground-floor lab built directly on solid building foundations, with the experimental setup positioned on top of a heavy ceramic bench-top, was found to provide sufficient passive control to minimize surface vibrations of the water bath.  Once the bath was set up and leveled, the sequence of images was projected and collected automatically, without moving any components.   As shown in Fig.~\ref{fig:3}(b), the undulating surface of the film distorted the pattern of dots, with the convex ridges acting like diverging lenses, and troughs like converging lenses.

The imaging process was repeated with and without the wrinkled film, resulting in 30 pairs of images, which were cropped to show a common region of interest that covered the floating film and some of the surrounding water surface.  These images were then thresholded and the pixel coordinates of the centroids of the dots were tabulated.  Connected regions with areas of less than 50 pixels were treated as noise, and excluded; the dots typically had areas of 250--400 pixels.  All image processing was performed in Matlab.

The initial ($\bm{x_i}$, without film) and final ($\bm{x_f}$, with film) locations of the dots were paired up using a simple particle-tracking algorithm, as demonstrated in Fig.~\ref{fig:3}(c).   A search radius of 0.07 cm was taken around each dot, and within this radius the closest possible pairings of dots between the initial and final images were made.  The apparent motion of each dot between the two images defined its displacement $\bm{\delta}=\bm{x_f}-\bm{x_i}$.  

By using an image sequence, with well-separated features in each image, this method can achieve an arbitrarily dense set of point pairings that sample the surface deformation, while allowing for a robust feature-matching algorithm within each image pair. For example, Fig.~\ref{fig:3}(d) shows a stacked image of 30 dot patterns, as seen through a wrinkled film. At this density, the dots allow for a good resolution of features as small as 1 mm.   

\begin{figure}[t]
\includegraphics[width=\linewidth]{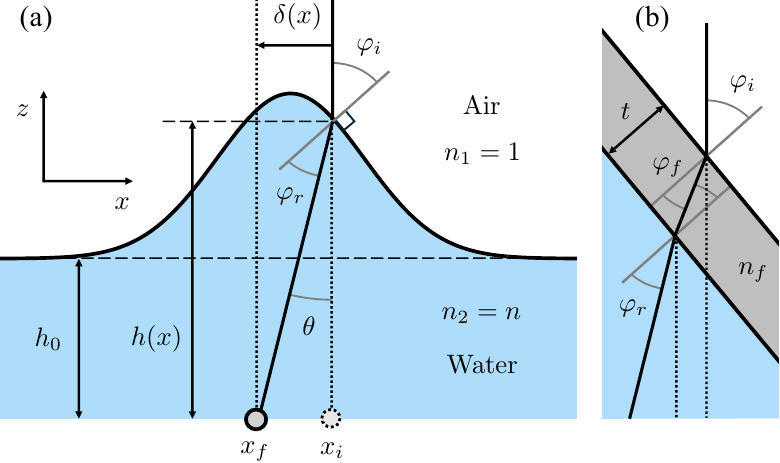}
\caption{\label{fig:4} Ray tracing diagrams showing how light rays are refracted through a water layer with an uneven upper surface.  (a) The apparent position of a dot is displaced by a distance $\delta$, as its image passes through that surface. (b) Close-up, the film can be considered as a thin surface layer of thickness $t$, with negligible effect on the displacement pattern if $t\ll h$.}
\end{figure}

\section{Synthetic schlieren imaging}
\label{Synthetic schlieren imaging}

The experimental setup used in this work defines an inverse optics problem: knowing how the wrinkled surface lenses light, we want to work out its shape.   A cross-sectional sketch of this problem, in the form of a ray-tracing diagram, is shown in Fig.~\ref{fig:4}(a).  In this section we describe our approach to solving this inverse problem and reconstructing the shape of the wrinkled film.  After validation of the method, we summarize how the films are analyzed to extract the wavelengths and amplitudes of the resulting patterns.

\subsection{Surface reconstruction} 
\label{surface reconstruction}

Consider a flat water bath of initial height $h_0$, with a set of point-like dots displayed at its base.  The addition of a thin film, of nominal thickness, causes the surface to adopt a deformed shape, with a new height $h$ that varies with position.  As viewed from above, a dot initially at position $x_i$ will now be seen at an apparent position $x_f$, displaced by a distance $\delta =x_f - x_i$.  
The ray traced from an overhead camera to this apparent position is refracted by an angle $\theta$ at the air-water interface.   For this, we assume that the surface height is single-valued (i.e. that it doesn't fold over itself) and that any ray intersects the surface only at a single point.  In the paraxial approximation (parallel rays, or a distant camera), the deformed height of the water layer at $x_i$ is
\begin{equation}
\label{eq:1}
h = \frac{\delta}{\tan \theta} = \frac{\delta}{\tan(\varphi_i - \varphi_r)},
\end{equation}
where $\varphi_i$ and $\varphi_r$ are the the angles of incidence and refraction that the ray makes to the air-water interface, respectively, and where $\theta=\varphi_i-\varphi_r$.  These angles are related by Snell's law, $ \sin{\varphi_i} = n\sin{\varphi_r}$, assuming refractive indices $n_1 = 1$ for air, and $n_2 = n$ for water.  As sketched in Fig.~\ref{fig:4}(b), the presence of a thin film of constant thickness $t$ and refractive index $n_f$ does not affect this relationship, as refraction at the air-film and film-water interfaces ensure that $\sin(\varphi_i) = n_f\sin(\varphi_f) = n\sin(\varphi_r)$, where $\varphi_f$ is the refracted angle of the ray in the film.  Using Snell's law to eliminate $\varphi_r$ then gives 
\begin{equation}
\label{eq:2}
    h = \frac{\delta}{\tan(\varphi_i - \sin^{-1}(n^{-1} \sin \varphi_i))},
\end{equation}
which, in the small angle approximation, simplifies to
\begin{equation}
\label{eq:3}
    h \approx \frac{\delta}{\varphi_i}\ \frac{n}{n-1}.
\end{equation}
This approximation is second-order accurate in $\varphi_i$, with a relative error below 1\% up to $\varphi_i = 0.20$, and below $5\%$ up to $\varphi_i = 0.45$.   By making the similar approximation that $\dv*{h}{x} = \tan\varphi_i \approx \varphi_i$ (also second-order accurate, relative error below $5\%$ up to $\varphi_i = 0.37$), and rearranging, we obtain a relationship between the observed displacement of the dot and the height of the deformed surface above it: 
 \begin{equation}
 \label{eq:4}
    \delta = \frac{n-1}{n}\ h\ \dv{h}{x} = \ \frac{n-1}{2n}\ \dv{}{x}{(h^2)}.
 \end{equation}
Since this derivation follows identically in any direction, it is easily generalized into two dimensions, where $\bm{\delta} = (\delta_x,\delta_y)$ is the apparent displacement vector of a dot in the $x$ and $y$ directions. In vector form, then,
 \begin{gather}
 \label{eq:5}
   \bm{\delta} = \frac{n-1}{2n} \ {\bm{\nabla}}{(h^2)}.
\end{gather}
 Finally, by taking the divergence of this equation, rearranging and simplifying, we obtain the Poisson equation
\begin{align}
\label{eq:6}
\nabla^2 \eta = \bm{\nabla \cdot \delta}
\end{align}
where $\eta = h^2 (n-1) / (2n)$.  Given appropriate boundary conditions, along with the locations and displacements of a large set of discrete points, Eq.~\ref{eq:6} can be efficiently and accurately solved for $\eta$, through use of an interpolating function for $\bm{\nabla \cdot \delta}$.  

Previous approaches to synthetic schlieren imaging have simplified and solved Eq.~\ref{eq:4} in the small-amplitude limit~\cite{moisy2009synthetic,gomit2022free}, by making the approximation that $\bm{\delta} \approx ((n-1)/n)h_0\bm{\nabla} h$.  The change of variables we use to develop Eq.~\ref{eq:6} instead leads to a linear system of equations that can be solved without assuming that surface deformations are small in amplitude, i.e. avoiding the constraint that $|h - h_0| \ll h_0$.  It also avoids the over-constrained nature of inverting Eq.~\ref{eq:5}, and allows access to standard solvers of the Poisson equation.  

Here, we used the \texttt{scatteredInterpolant} function in Matlab to interpolate smoothly between the displacement data of the imaged dot locations.  Taking $n = 1.33$ for the water bath, and in the case where the water surface retains its initial height of $h_0= 5\,$ mm far from the film, we used \texttt{solvepde} to solve the resulting inverse problem for $\eta$ and hence for the height, $h(\mathbf{x})$, of the distorted film surface.  Finally, we note that, as long as the film is thin compared to the water bath, $t\ll h$, the different refractive index of the film has only a minimal effect on these calculations, and so is ignored here for simplicity.  For example, calculating ray deflections using the three-layer model sketched in Fig.~\ref{fig:4}(b) would lead to corrections of less than $10\,\mu$m in $\delta$, for angles up to $\varphi_i = 1.0$ and assuming a thickness of $t= 100\,\mu$m and refractive index of $n_f = 1.40$ for the PDMS film.

\subsection{Validation and feature extraction}
\label{Validation and feature extraction}

The quality and resolution of the surface reconstruction depends on the area density, $\rho = N/a$, of $N$ dots scattered over an imaged area $a$.  For features with a characteristic size of $\lambda$, a lower bound on the effective resolution is set by the Nyquist condition, $\rho \lambda^2 \geq 4$, corresponding to a sampling interval of half the feature size, on average.  

\begin{figure}[t]
\includegraphics[width=\linewidth]{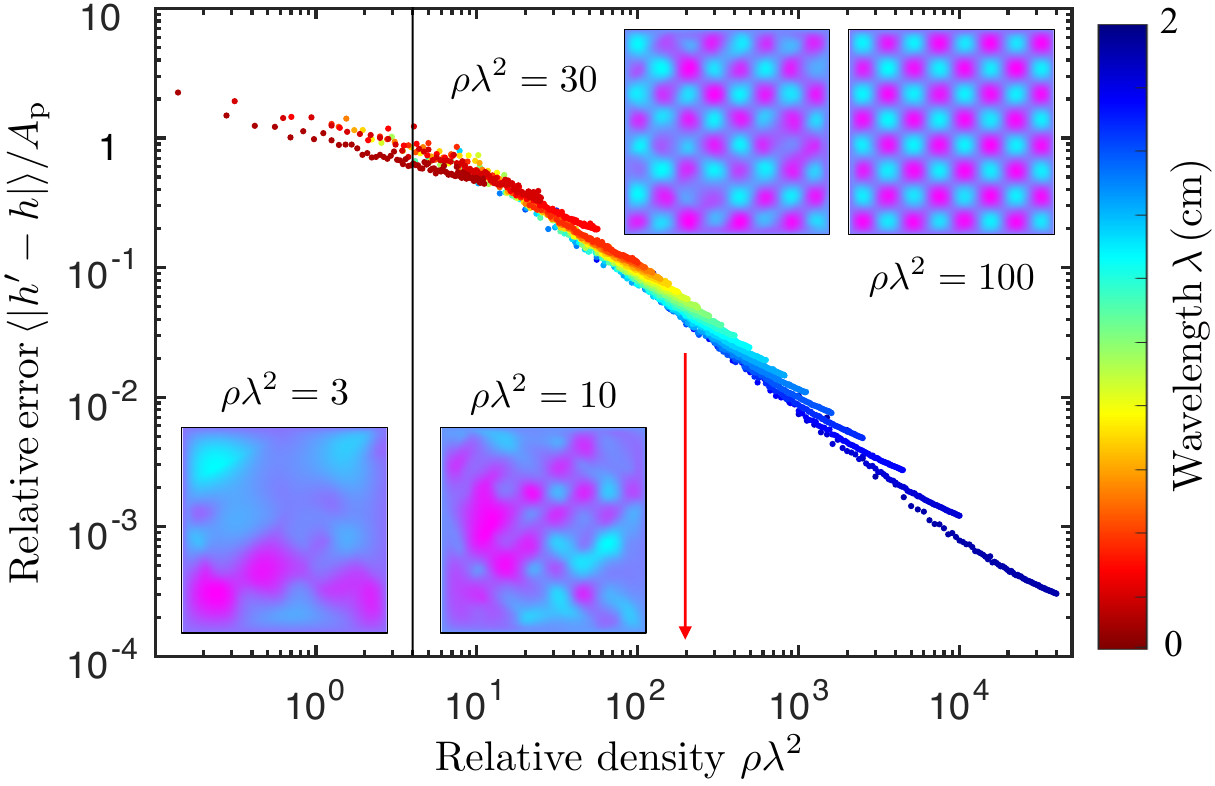}
\caption{\label{fig:5} Surface reconstruction methods were tested with artificial data from a sinusoidal surface.  Insets show results for surfaces of width $W=4$ cm and wavelength $\lambda = 1$ cm.  The accuracy of the reconstructed surface depends on the relative density of dots used, $\rho\lambda^2$.  Above the Nyquist condition (solid line, $\rho\lambda^2 = 4$), the relative error improves as 1/$\rho\lambda$.  Our experimental conditions were around $\rho\lambda\simeq 200$ (red arrow).}
\end{figure}

To validate and to test the resolving power of our methods, we used artificial input data, corresponding to the exact displacements expected for $N$ points randomly sampling a surface with height $h = h_0 + A_\mathrm{p} \sin{(2\pi x/\lambda)}\sin{(2\pi y/\lambda)}$.  To ensure the small-angle limit, we used a nearly flat surface with a peak amplitude of $A_\mathrm{p} = 2 \,\mu$m.  The data for $\bm{\delta}$ were then processed as described in Section~\ref{surface reconstruction}, and the height map of the reconstructed surface, $h^\prime$, was compared to its expected shape, $h$.  As shown in Fig.~\ref{fig:5}, the relative error of the reconstruction, $\langle | h^\prime - h|\rangle / A_\mathrm{p}$, was of order one below the Nyquist cutoff, and decreased with a scaling of $1/\rho \lambda^2$ as the dot density increased.  Our experiments were typically done under conditions where $\rho \lambda^2\simeq 200$.

\begin{figure}[t]
\includegraphics[width=1\linewidth]{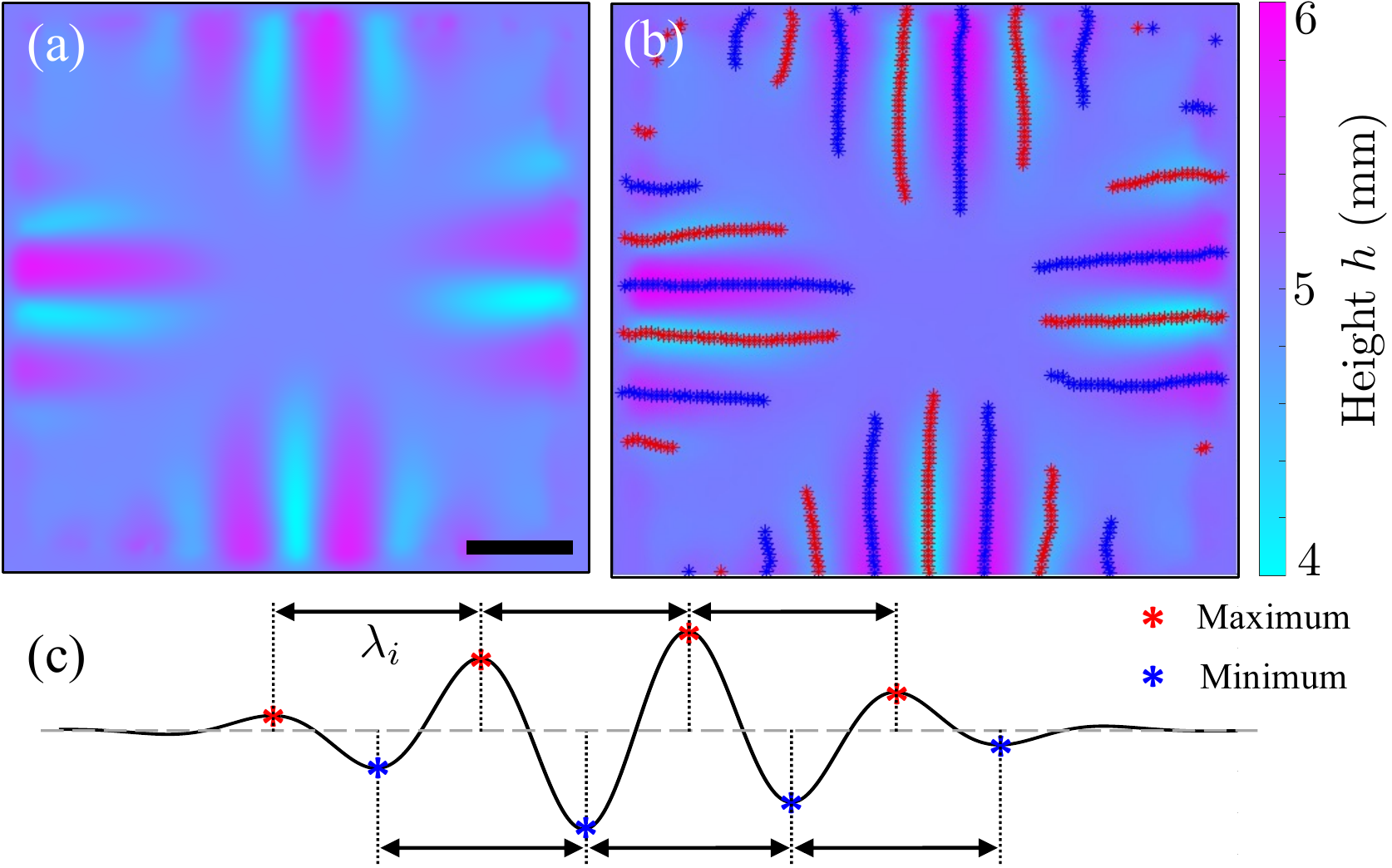}
\caption{\label{fig:6} Surface reconstruction and feature detection.  (a) The reconstructed surface height map for a typical wrinkled film is shown, with a negative Gaussian curvature defined by the two radii of curvature $R_1 = - R_2  =$ 70 mm.  The scale bar is 1 cm.  (b)  The locations of the maxima (red) and minima (blue) of the height $h$ are shown, superimposed over the film surface.  (c) These extrema are determined through cross-sectional profiles, on lines parallel to the film edges, and are used to measure the average wavelength $\lambda$ of the wrinkles.}
\end{figure}

From our experimental surface height maps, like the one shown in Fig.~\ref{fig:6}(a), we measured the wavelength and amplitude of the emergent wrinkles. Each film covered a square cross-sectional area of width $W = 50$ mm, and its surface was divided into a series of line segments parallel to the edges of the film. Surface height profiles were extracted along the lines, and the local minima and maxima of these profiles were found using Matlab's peakfinding algorithm, enforcing a minimum prominence of 0.2 mm. A typical result of this process is given in Fig.~\ref{fig:6}(b).  Wrinkle wavelengths were then measured between all sequential pairs of minima or maxima along each line, as in Fig.~\ref{fig:6}(c), and averaged.  For amplitude, the root-mean-squared (RMS) amplitude of the entire wrinkle pattern was calculated using the height variations across the whole film, relative to the water bath surface, $A = \sqrt{\langle|h-h_0|^2\rangle}$.  Using height profiles along the line segments parallel to the film edges, we also measured how the RMS amplitude varied with position away from the edge or center of the films.  

The experimental imaging methods were validated using a polyethylene terephthalate (PET) sheet with a fixed, rigid wrinkle structure. The same sheet was imaged under our synthetic schlieren methods and with optical coherence tomography (OCT)~\cite{manukyan2013imaging, cheung2015ultra}.  The OCT data was analyzed using ImageJ~\cite{ImageJ} to determine the peak amplitude, $A=0.32 \pm 0.03$ mm, and wavelength, $\lambda=5.4 \pm 0.2$ mm, of the wrinkles.
These results agree within error to the values determined using our methodology, of $A = 0.33 \pm 0.07$ mm and $\lambda = 5.7 \pm 0.2$ mm.

Finally, to evaluate the noise levels introduced by any vibrations, we compared two sets of dot patterns taken through a water bath at different times, in the absence of any floating film.  The average displacement of these dots showed a random distribution, with an average value of $\langle|\bm{\delta}|\rangle=23\,\mu$m, close to the 19 $\mu$m pixel resolution of the camera used.  The surface reconstructed from these displacements was flat, with a root-mean-squared roughness of 29 $\mu$m.

\section{Theory of wrinkling shells}
\label{Theory of wrinkling shells}

A theoretical framework for parameterizing thin floating shells has recently been developed~\cite{Aharoni2017,Albarran2018curvature,Tobasco2022,Tobasco2021}, which we adopt here.  The film's geometry is captured by its lateral size $W$ and Gaussian curvature, $\kappa$.  The principal curvatures $\kappa_1 = 1/R_1$ and $\kappa_2 = 1/R_2$ are expected to affect the morphology of the film only through $\kappa = \kappa_1 \kappa_2$, as a result of Gauss' \textit{theorema egregium}~\cite{Aharoni2017}.

There are four additional parameters predicted to affect the wrinkling pattern of a floating elastic shell.  These are all related to how energy can be distributed within the system~\cite{Tobasco2021}. The stretching modulus, $Y$, describes how resistant the material is to elastic deformation when an external force is applied.  It characterizes the energetic cost of changing the surface area of the film and is related to its Young's modulus, $E = 2.2$ MPa, and thickness, $t\simeq 100$ $\mu$m, by $Y = Et$.  The bending modulus, $B$, gives the relative difficulty of local bending or flexing of the film. For a film of Poisson ratio $\nu=1/2$, $B = Et^3/[12(1-\nu^2)] = Et^3/9$.  The weight of the water displaced by the film is accounted for by a gravitational stiffness, $K_g = \rho g$, where $\rho=1000$ kg m$^{-3}$ is the fluid density and $g = 9.81$ m s$^{-2}$ is the magnitude of the acceleration due to gravity.  Finally, the surface tension $\gamma$ characterizes the forces acting on the edges of the film, from the surrounding water bath.

\begin{table} [t]
\caption{\label{tab:table1}
Characteristic parameters of the wrinkled films.}
\begin{ruledtabular}
\begin{tabular}{lcc}
\multicolumn{1}{c}{\textrm{Parameter}}&
\textrm{Symbol}&\textrm{Value or Range}\\
\colrule
Domain size & $W$ & 5$\cdot $10$^{-2}$ m \\
Gaussian curvature & $\kappa$ & $|\kappa |$ = (0.1--1.1)$\cdot $10$^{3}$ m$^{-2}$ \\
Stretching modulus & $Y$ & (2.2--2.6)$\cdot$10$^2$ N m$^{-1}$ \\
Bending stiffness & B & (2.4--4.0) $\cdot$10$^{-7}$ N m \\
Gravitational stiffness & K$_g$ & 9.8$\cdot $10$^3$ N m$^{-3}$ \\
Surface tension & $\gamma$ & 7.2$\cdot $10$^{-2}$ N m$^{-1}$ \\
\colrule
Bendability  & $YW^4|\kappa|/B$ & (0.5--5)$\cdot $10$^{6}$ \\
Deformability  & $Y|\kappa|/K_g$ & (3--27) \\
Confinement  & $YW^2|\kappa|/\gamma$ & (1--9)$\cdot $10$^{3}$ \\
\end{tabular}
\end{ruledtabular}
\end{table}

Ratios of the characteristic energies involved in stretching, bending, vertical displacement, and surface tension provide a useful set of dimensionless groups for studying wrinkle patterns.  Following the definitions of Tobasco~\cite{Tobasco2021}, our films have high bendability, $YW^4|\kappa|/B\sim 10^{6}$, moderate substrate deformability (or specific thickness~\cite{Albarran2018curvature}), $Y|\kappa|/K_g \sim 10$ and weak surface tension (or strong geometric confinement), $YW^2|\kappa|/\gamma\sim 10^{3}$.   The full range of the parameter space explored in our experiments is given in Table~\ref{tab:table1}, and values for individual films are collated in Table S1 of the online supplemental information~\cite{SInote}.  In practical terms, this parameter range means that surface tension will be negligible, as has been demonstrated with experiments involving even thinner PDMS shells~\cite{Aharoni2017}.   By balancing the bending stiffness $B$ with the substrate stiffness $K_g$, the wavelength of any emergent wrinkles is predicted~\cite{Finn2019,Paulsen2016,cerda2003geometry} to be
\begin{align}
 \label{eq:7}
\lambda = 2 \pi \left(\frac{E t^{3}}{9 K_g}\right)^{1/4} =  2 \pi \left(\frac{B}{K_g}\right)^{1/4} = 2 \pi \Lambda.
\end{align}
Here, $\Lambda = (B/K_g)^{1/4}$ is a characteristic scale of features for which the bending energy of the film and the gravitational energy of the substrate will be roughly equal.  It is about 2.4 mm for our films.

The characteristic amplitude of the wrinkled film follows from its need to accommodate the additional surface area of its naturally curved surface, compared to the cross-sectional area that it occupies on the water bath.  In other words, the shell's true surface area $\mathcal{A}$ exceeds that of its projected surface area $W^2$, and this excess area is expected to remain within its deformed or wrinkled surface~\cite{Aharoni2017}.   Given the connection between the Gaussian curvature and the metric of a surface~\cite{Berger2003}, in their undeformed state our curved films will have a surface area
\begin{equation}
\label{eq:8}
\mathcal{A} = W^2 \big(1+\frac{|\kappa| W^2}{12} +  O(\kappa^2 W^4)\big).
\end{equation}
Compared to their flat projection, these films can be described as having a relative excess of area of $|\kappa| W^2/12$, to leading order.  A comparative case of a deformed surface could be a regular array of wrinkles of RMS amplitude $A$ (i.e. peak amplitude of $A_\mathrm{p} = \sqrt{2}A$) and wavelength $\lambda$.  The surface area of such a pattern,
\begin{equation}   
   \label{eq:9}
   \mathcal{A} = W^2  \big(1 + \frac{2\pi^2 A^2}{\lambda^2}+  O(A^4/\lambda^4)\big),  
\end{equation}
can be found by integrating over the path length of its sinusoidal cross-section.  Comparing the relative excess of area in Eqs.~\ref{eq:8} and \ref{eq:9} leads to the prediction that ${2\pi^2 A^2}/{\lambda^2}\simeq{|\kappa| W^2}/{12}$, assuming that surface area is conserved. While our wrinkle patterns are more complex than this test case, the scaling relationship, $A^2/\lambda^2 \sim|\kappa| W^2$, should remain generally valid, in the absence of any significant stretching of the film.

\begin{figure}[t!]
\includegraphics[width=\linewidth]{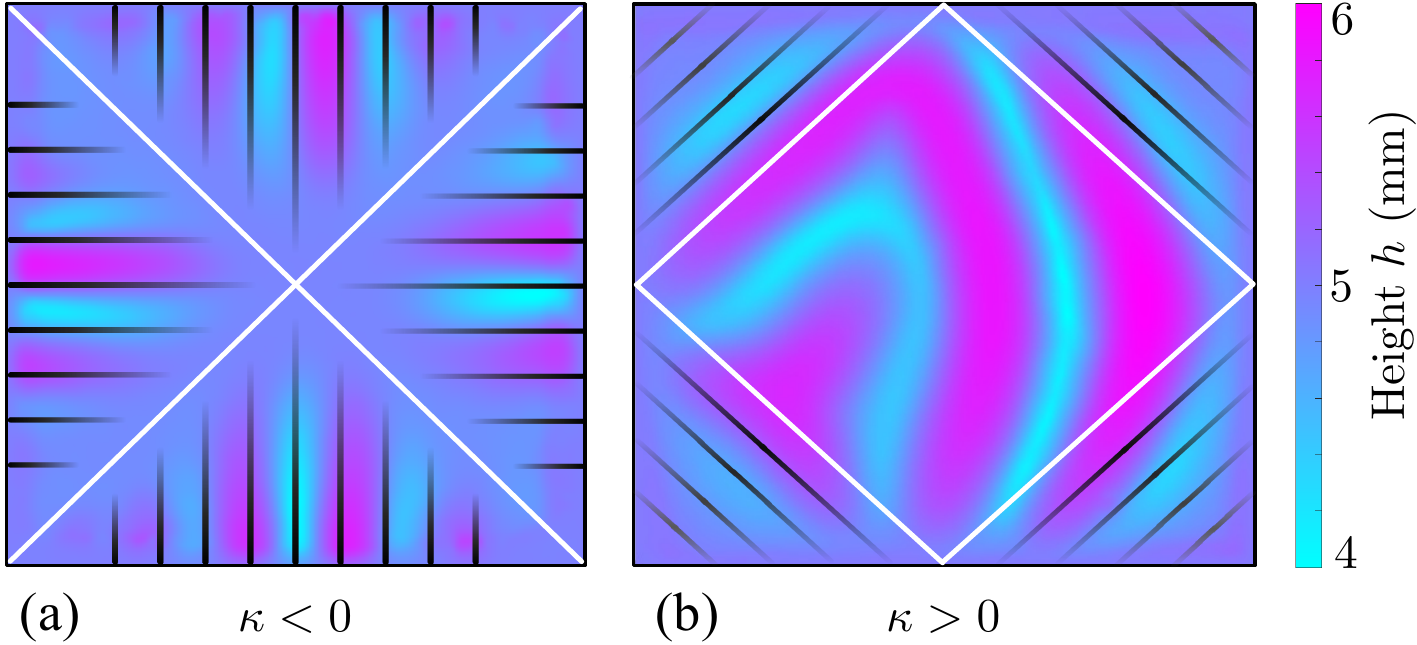}
\caption{\label{fig:7} Stable lines define the ordered regions expected on wrinkling curved sheets; the black lines in both panels show the predicted domains of ordered wrinkles, along with their orientation.  (a) For our negatively curved surfaces, the wrinkle amplitude is anticipated to decay as one moves towards the medial axes, shown in white, and away from the shell boundaries. (b) Disordered wrinkles are predicted to occur in the central area of the positively curved surfaces, with the wrinkle amplitude decaying in the ordered corner domains as one moves towards the boundary.}
\end{figure}

Finally, for floating curved sheets, wrinkles have been shown to arrange into ordered domains, each of which can have wrinkles aligned in a different direction~\cite{Aharoni2017}.  The structure of the domains, and the alignment of the wrinkles within them, can be predicted from the shape of the boundary of the shell, and the sign of the Gaussian curvature, as sketched in Fig.~\ref{eq:7}.  These predictions follow the method of stable lines, developed by Tobasco~\cite{Tobasco2021} and collaborators~\cite{Tobasco2022}.  For negatively curved shells, the domain boundaries are expected to follow the medial axis, the set of points which are equally close to more than one location along the boundary, as in Fig.~\ref{eq:7}(a).  For positively curved shells, the corresponding domain boundaries are given by a reciprocal figure construction~\cite{Tobasco2022,maxwell1864xlv}.  For the case of square boundary to the shell area, this predicts triangular domains of wrinkles in the corners of the film, with a disordered central domain, as in Fig.~\ref{eq:7}(b).

\section{Results and Discussion} 
\label{Results}

\begin{figure*}[t!]
\includegraphics[width=0.95\linewidth]{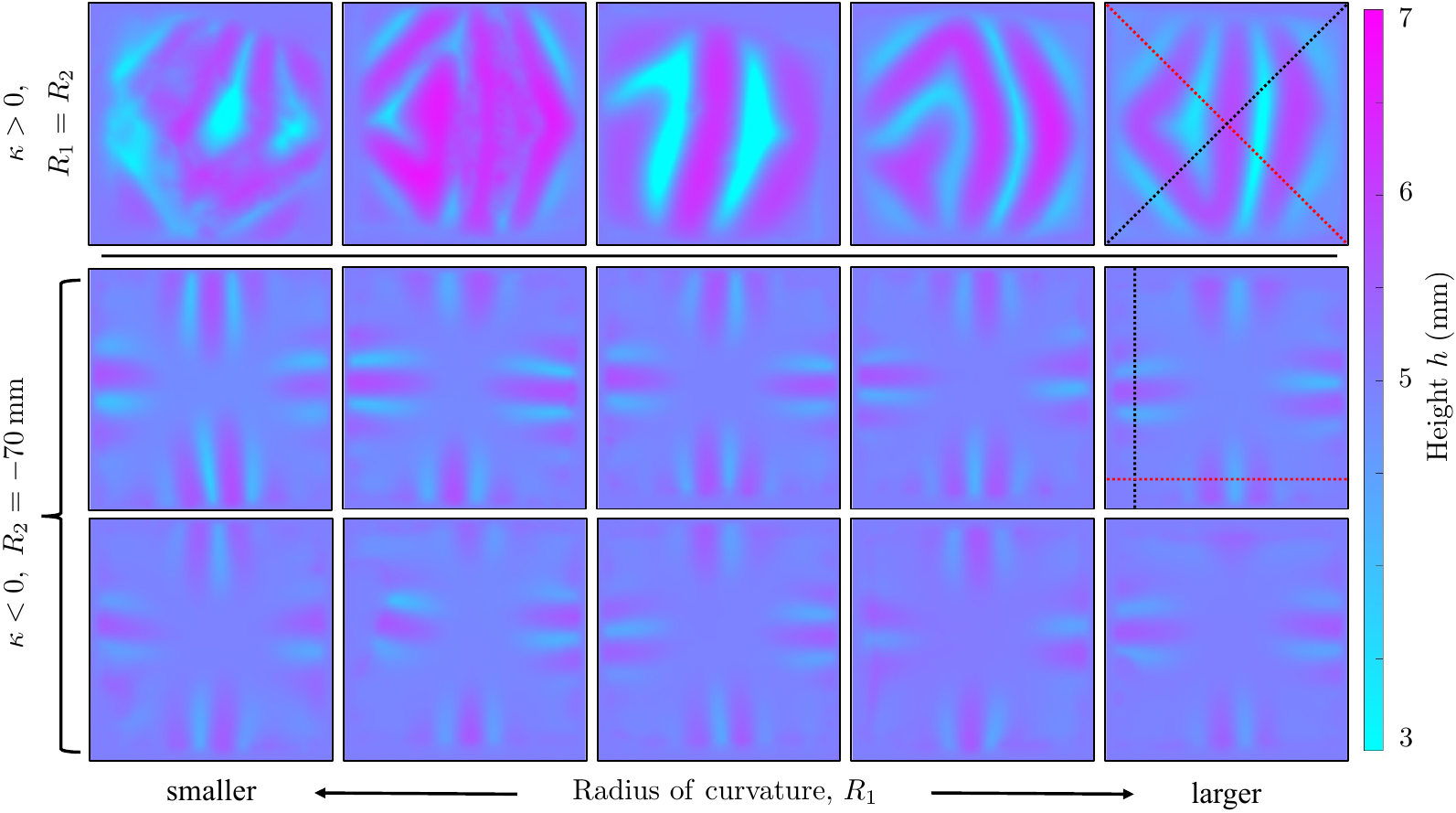}
\caption{\label{fig:8} Reconstructed surface maps for floating curved sheets of various inherent Gaussian curvatures $\kappa = 1/R_1R_2$. The top row shows films cut from spherical surfaces with radii of $R_1 = R_2 =$ 30, 40, 50, 60 and 70 mm.  Films with negative Gaussian curvature were cut from saddle-shaped surfaces with different $R_1$, and fixed $R_2 = -70$ mm.  The middle row shows the wrinkle patterns for $R_1 =$ 40, 50, 60, 70, and 80 mm, while the bottom row shows the patterns for $R_1 =$ 90, 100, 110, 120 and 130 mm.  In the right, upper two panels, dashed lines indicate the locations of the cross-sectional profiles shown in Fig.~\ref{fig:9}}
\end{figure*}

\begin{figure}
\includegraphics[width=\linewidth]{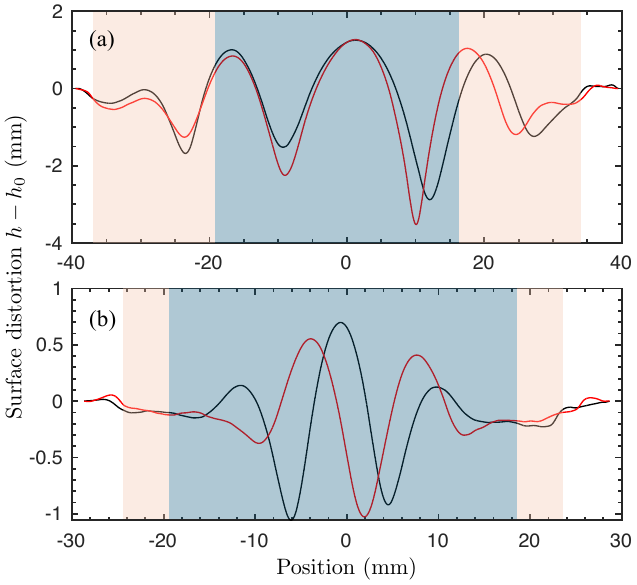}
\caption{\label{fig:9} Cross-sectional profiles of floating shells with (a) positive and (b) negative intrinsic curvature.  Profiles are extracted at the locations shown by the dashed lines in Fig.~\ref{fig:8}.  Red and blue shading highlight locations that belong to different domains, as predicted by the method of stable lines (see Fig.~\ref{fig:7}).}
\end{figure}

The deformed shapes of curved floating sheets with fifteen distinct geometries were measured using the synthetic schlieren imaging methods developed in Section~\ref{Synthetic schlieren imaging}, and used to test recent predictions of how such shells wrinkle.  To aid in this comparison, the dimensionless bendabilty, deformability and confinement were calculated for each film (see Table S1 in the online supplementary information~\cite{SInote}), based on their size $W$, Gaussian curvature $\kappa$ and thickness $t$.  We focus on the wrinkling regime of these experiments, where the effective confinement is not strong enough to generate folding. 

Wrinkling domains from representative experiments with positive and negative Gaussian curvature are compared with the predictions of the method of stable lines~\cite{Tobasco2022,Tobasco2021} in Fig.~\ref{fig:7}.  The medial axis of the negatively curved shells, $\kappa<0$, accurately describes the division of the wrinkled films into four triangular domains, with wrinkles perpendicular to the outer edges of the film.  For positively curved shells, the reciprocal figure construction outlines an interior region of the film in which a relatively disordered, high-amplitude wrinkle pattern appears.  Wrinkles in the four corners of the film are also oriented at 45$^\circ$ to the film's edges, as predicted; these more ordered domains are comparatively small here, only spanning one or two wavelengths across. 

Figure~\ref{fig:8} gives the reconstructed surface profiles of all the films used in this study, with some cross-sectional profiles across films also given in Fig.~\ref{fig:9}. As can be seen for the full set of experiments, our representative descriptions of the wrinkled domains hold for shells with a wide range of Gaussian curvatures.  In general, we find that varying the magnitude of $\kappa$ only changes the amplitude of the resulting pattern, without otherwise affecting its structure.  The films with positive curvature, $\kappa > 0$, all show a disordered central region, with a few diagonal wrinkles appearing in the corners of the film. The films with negative curvature, $\kappa < 0$, all show flat regions along the stable lines, separating four distinct domains of well-ordered wrinkles.  These consistent features allow us to develop robust measures of wrinkle wavelength and amplitude, extracted from the surface profile data.

These results also demonstrate that the wrinkles depend only on the Gaussian curvature, rather than on the individual values of the principal curvatures $\kappa_1$ and $\kappa_2$.  For the negatively curved films shown in the lower two rows of Fig.~\ref{fig:8}, these are the curvatures that would be measured along the perpendicular edges of a film, and can differ from each other by up to about a factor of two.  There are no corresponding differences in the wavelength, amplitude, or positioning of the wrinkles that appear along these directions, however, and no evidence of symmetry-breaking due to the differences in principal curvatures.  Although expected from Gauss' \textit{theorema egregium}~\cite{Aharoni2017,Berger2003}, this remains an unintuitive and remarkable result. 

\begin{figure}[t]
\includegraphics[width=\linewidth]{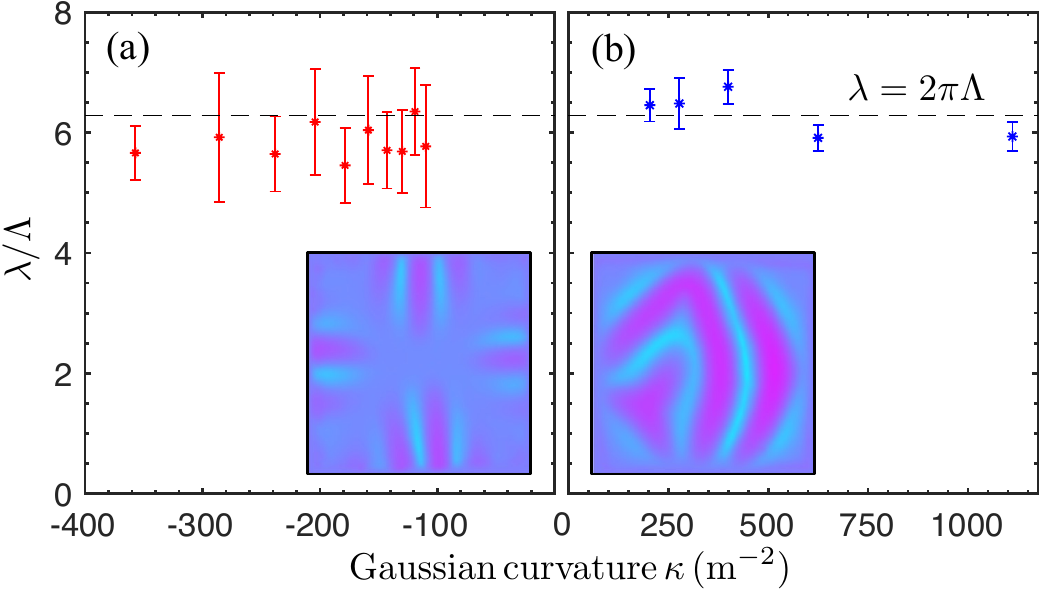}
\caption{\label{fig:10} The characteristic wavelength of the wrinkling of thin curved films agrees with the predictions of elastica-based models~\cite{Finn2019,Paulsen2016,cerda2003geometry}.  Wavelengths $\lambda$ are measured for floating films with (a) negative and (b) positive Gaussian curvature $\kappa$, as the average distance between successive troughs or peaks of the surface $h$, measured on lines parallel to the film's edges.  The dashed lines show the prediction that $\lambda = 2\pi \Lambda$. }
\end{figure}

The wavelength of a wrinkled film is measured here as the average distance between consecutive peaks or troughs of the wrinkles, and represents the spatial periodicity of the pattern.  As described in Section~\ref{Validation and feature extraction}, we made use of the regular domain structure to identify the peaks and troughs of the reconstructed film surface, and to find the separation of these features along lines parallel to the film edges.  Based on a balance of the bending energy of the film and the gravitational energy of displacing the fluid substrate~\cite{Finn2019,Paulsen2016,cerda2003geometry}, Eq.~\ref{eq:7} predicts a wrinkle wavelength of $\lambda = 2\pi\Lambda$, where the natural length scale $\Lambda = (B/K_g)^{1/4}$ is 2.4$\pm$0.1 mm.  As shown in Fig.~\ref{fig:10}, the measured wrinkle wavelengths are in good agreement with this prediction.  Specifically, the ratio $\lambda/\Lambda$ remains unaffected by both the sign and magnitude of $\kappa$, as can be seen by comparing the results for negative curvatures in Fig.~\ref{fig:10}(a) and positive curvatures in Fig.~\ref{fig:10}(b).

\begin{figure*}[t!]
\includegraphics[width=\linewidth]{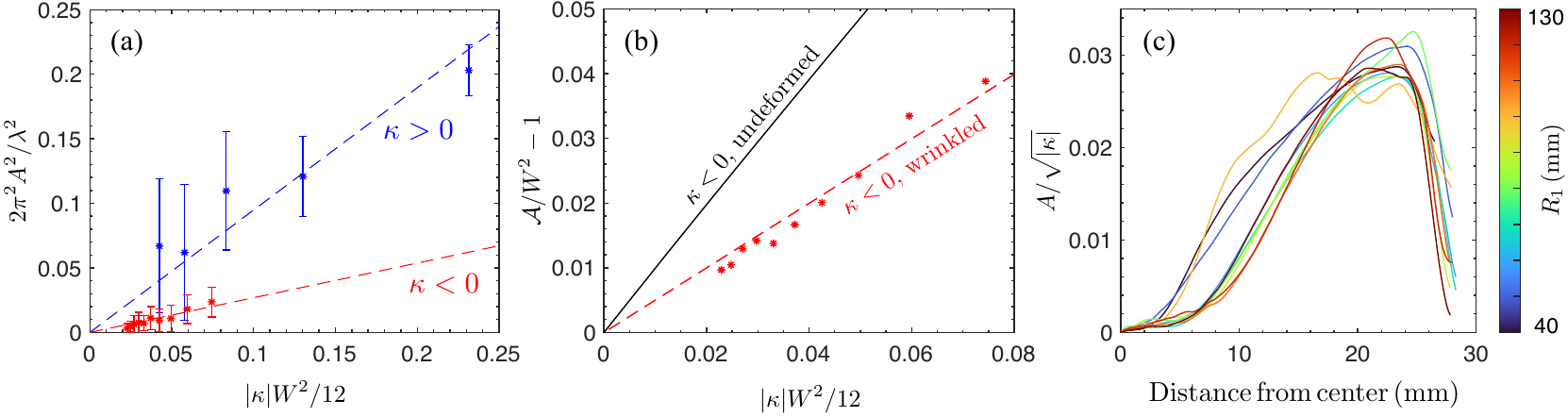}
\caption{\label{fig:11} Wrinkle amplitudes are related to a film's natural Gaussian curvature, and to the excess of its surface area as compared to a flat cross-sectional profile. (a) The relative excess area, as estimated from a wrinkled film's RMS amplitude, $2\pi^2A^2/\lambda^2$, can be compared to the excess area of the undeformed shell, $|\kappa| W^2/12$.  While these measures are proportional to each other, the sign of $\kappa$ has an effect on how the area is distributed: the spherical shells (blue) show larger amplitudes and areas, compared to the hyperboloid shells (red).  (b) For the negatively curved shells, systematically measuring the total surface area, $\mathcal{A}$, of the wrinkled and undeformed configurations shows that there is a deficit in relative excess area, $\mathcal{A}/W^2-1$, for the wrinkled films. (c) For the negatively curved shells, the wrinkle amplitude increases with distance away from the center of the film, and with $\sqrt{\kappa}$.  There are, however, flat regions of width $\sim 5$ mm around the film center, and the stable lines.}
\end{figure*}

The amplitude of the wrinkle patterns, and the excess surface area that these features accommodate, are explored in Fig.~\ref{fig:11}.  As an initial measure of the amplitude of a pattern we use the RMS amplitude of the surface distortion, $A = \sqrt{\langle|h-h_0|^2\rangle}$, measured over the entire film.  If the wrinkles had a uniform sinusoidal cross-section of wavelength $\lambda$ and amplitude $A$, then they would have an excess surface area of $2\pi^2A^2/\lambda^2$, relative to a flat film with the same cross-section.  The undeformed shell will also have a relative excess of area, given by $|\kappa|W^2/12$.  These measures both represent a type of misfit strain between the natural surface area of the film, and the cross-sectional area of the fluid bath that it covers.  They are derived from Eqs.~\ref{eq:8} and~\ref{eq:9}, and compared to each other in Fig.~\ref{fig:11}(a).  For the positively curved films, where the assumption of a constant amplitude is more plausible (see top row of Fig.~\ref{fig:8}) a least-squares regression gives a slope of 0.95$\pm$0.21.  This suggests that the film's surface area is reasonably well conserved as it wrinkles, in line with the assumptions of most models~\cite{Aharoni2017,Tobasco2021,Tobasco2022}.  For the negatively curved shells, however, the best-fit line between these two measures of excess area has a slope of only $0.27\pm0.04$.  This is a clear difference in how the sign of the curvature affects the wrinkle amplitude.  

For the negatively curved shells, the wrinkling amplitude varies considerably over the film, with flat areas surrounding the stable lines.  To more carefully evaluate these shells, we used the \texttt{surfarea} function~\cite{Surfarea} in Matlab to measure the total surface area of the reconstructed height maps of their wrinkled surfaces.  From these areas $\mathcal{A}$ we directly calculated an excess area, $\mathcal{A}/W^2 - 1$, relative to the area of the water which they cover, $W^2$.  We used the same methods to measure the relative excess area of hyperboloid surfaces with constant curvatures, for comparison, with the same lateral dimensions as our molds. As shown in Fig.~\ref{fig:11}(b), the wrinkled films made from negatively curved surfaces have considerably less surface area than the reference, undeformed shells.  Specifically, while the reference shells show good agreement with the expected limit of $\mathcal{A}/W^2 -1 = |\kappa|W^2/12$, these two measures of the relative excess area differ by a best-fit factor of $0.50\pm0.04$ for the wrinkled configurations.   In other words, these floating curved sheets have lost about half of their excess area.  

A potential explanation for the changes to the films' surface areas during wrinkling lies in the relatively low substrate deformability of these experiments.  Although bending is by far the most energetically favorable mode of deformation, the energy costs associated with substrate deformation and film stretching differ by only about a factor of ten (see Table~\ref{tab:table1}).  Some stretching or compression, with an associated change to the surface area, could thus reasonably accompany the surface wrinkling.  The sign of the Gaussian curvature would contribute to this effect: the spherical cutouts have a deficit of perimeter compared to the flat films, while the hyperbolic cutouts have an excess of perimeter.  In order to accommodate these shapes on a flat surface, the surfaces with positive Gaussian curvature need to be stretched out, while the negatively curved shells would need to be compressed.

Additional evidence of some film compression, for the negatively-curved shells, can be seen in Fig.~\ref{fig:11}(c).  This plot shows how the RMS amplitude varies from the center of these films, to their edges.  The averages are calculated on concentric square boundaries, along lines a fixed distance from the film's center.  Over much of the film the wrinkle amplitude increases smoothly with the distance from the center (practically, the distance from a stable line).  However, there is a flattened region of width $\sim 5$ mm near the center of the film, or around the stable lines, which is not predicted by the current models~\cite{Tobasco2021,Tobasco2022}.  In these regions, with vanishingly small wrinkle amplitude, local film compression would appear to be required.  

\section{Summary and Conclusions}

We investigated how the natural curvature of a film or sheet affects its wrinkling behavior, using a simple yet powerful non-invasive imaging technique. When curved films are floated on a fluid substrate they tend to wrinkle, due to a mismatch between the natural surface area of the film, and the space to which they are confined.  We found that the wavelength of their wrinkles has the same value as for flat confined sheets~\cite{cerda2003geometry,Finn2019}, regardless of the sign or magnitude of the film's curvature.  This represents a simple balance between the bending energy in the film and the energy cost of deforming the substrate.  Additionally, we showed that the amplitude of the wrinkles is related to the magnitude of the Gaussian curvature, by $A\sim \sqrt{\kappa}$.  This reflects how the excess area of the curved film is redistributed into the wrinkled surface.  Differences in the amplitudes of wrinkling between positively and negatively curved sheets suggest that stretching can remain a part of this process, and imply that assumptions about the conservation of area during wrinkling~\cite{Aharoni2017} may need to be treated cautiously.  Finally, the wrinkled films organize into different domains, each exhibiting wrinkles oriented in different directions. The structure of these domains is well-predicted by the method of stable lines, based on a film's shape and Gaussian curvature~\cite{Tobasco2021,Tobasco2022}.   However, we found that flat regions, with a width of a few millimeters, surrounded the stable lines.  These local absences of wrinkling are also consistent with some stretching effects.  

To achieve these results, we developed an imaging technique based on the methods of synthetic schlieren imaging, which have previously proved useful for generating topographical maps of the height $h$ of deformed liquid surfaces~\cite{wildeman2018real,moisy2009synthetic}, including for cases of a floating annulus~\cite{pineirua2013capillary} or drying droplets~\cite{Kilbride2023}.  This approach relies on measurements of the refraction of rays as they pass through a deformed, wrinkled surface.

Our approach does not require any assumption of small surface perturbations, i.e. it avoids the constraint that $(h - h_0) \ll h_0$, where $h_0$ is the reference height of the imaged surface.  The change of variables involved in this approach also circumvents the issue of over-constrained data encountered when using two scalar fields, namely the displacement of visual features in two independent directions, to reconstruct the single scalar field of the surface height.  We further demonstrated how the use of a series of projected features can increase the accuracy of the reconstructed surface, without introducing ambiguities in feature matching, and provided estimates of this accuracy, based on the density of sampling points and the Nyquist sampling condition. 

Together, these results provide insight into the effects of curvature on wrinkling phenomena, and an imaging method with broad applicability.  Since the stiffness introduced by a buoyant substrate is simply a linear restoring force, these findings are also appropriate to rigid films attached to soft elastic substrates, where there is some mismatch in curvature between the film and substrate;  This could provide for applications in shape memory, for example.   As PDMS is photoelastic~\cite{Tarjanyi2014}, like many other polymers, crossed-polarizers could be used in parallel with these methods, to evaluate local strains and stretching.   Local stretching could also be studied through detailed analysis of the perimeter to area ratio of circles inscribed within the film, either physically or by tracing the appropriate paths within the surface of a reconstructed film.  Other relevant curved films and interfaces are encountered in diverse settings, including in the brain~\cite{jalil2015cortical} and gut~\cite{Shyer2013}, swelling hydrogels~\cite{bertrand2016dynamics}, liquid lenses~\cite{Kilbride2023}, deformable~\cite{Ohzono2009} or reconfigurable~\cite{Paratore2022} microfluidics, stretchable~\cite{Rogers2010} and reconfigurable~\cite{Tian2021} electronics, flexible optical components such as solar panels and diffraction gratings~\cite{schauer2018disordered}, and graphene sheets~\cite{nano13010095}.  

\begin{acknowledgments}
We thank Mark Wilkinson for insight into solving the inverse optics problem, and in particular for suggesting the change of variables used here.  We are grateful to Kartikeya Walia for assistance with 3D printing the curved substrates, and to Haida Liang and Sammy Cheung for guidance and access to OCT instrumentation in the ISAAC lab at Nottingham Trent University.

\end{acknowledgments}



\end{document}